\documentclass[paper,twocolumn,showpacs,superscriptaddress,10pt]{revtex4-1} 
\usepackage{graphicx}
\usepackage{amsmath}
\usepackage{SIunits}
\begin{document}

\title{Frequency transfer via a two-way optical phase comparison on a multiplexed fiber network}

\author{C. E. Calosso$^{1}$, E. Bertacco$^{1}$, D. Calonico$^{1}$, C. Clivati$^{*,1,2}$, \\G. A. Costanzo$^{1,2}$, M. Frittelli$^{1,2}$, F. Levi$^{1}$,  A. Mura$^{1}$, A. Godone}

\address{
Istituto Nazionale di Ricerca Metrologica INRIM, strada delle Cacce 91, 10135, Torino, Italy
\\
$^2$Politecnico di Torino, Corso Duca degli Abruzzi 24,10129, Torino, Italy \\
*Corresponding author: c.clivati@inrim.it
}

\begin{abstract}We performed a two-way remote optical phase comparison on optical fiber. Two optical frequency signals were launched in opposite directions in an optical fiber and their phases were simultaneously measured at the other end. In this technique, the fiber noise is passively cancelled, and we compared two optical frequencies at the ultimate \(10^{-21} \) stability level.  The experiment was performed on a \unit{47}{\kilo\metre} fiber that is part of the metropolitan network for Internet traffic. The technique relies on the synchronous measurement of the optical phases at the two ends of the link, that is here performed by digital electronics. This scheme offers some advantages with respect to active noise cancellation schemes, as the light travels only once in the fiber.
\end{abstract}


\maketitle 

\begin{figure}[h!]
\centerline{\includegraphics[width=.8\columnwidth]{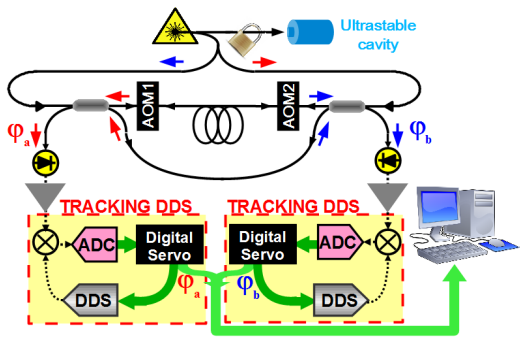}}
\caption{The optical apparatus and the electronic system: AOM Acousto Optic Modulators, ADC Analog to Digital Converters, DDS Direct Digital Synthesizers.}
\label{fig:setup}
\end{figure}

\begin{figure}[h!]
\centerline{\includegraphics[width=.8\columnwidth]{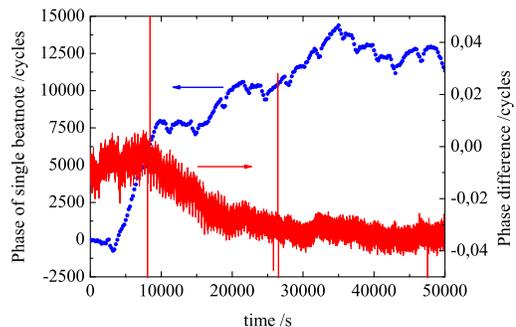}}
\caption{The phase  of one of the two independent beatnotes (left-hand axis, blue circles) and of their difference (right-hand axis, red line). 1 cycle is about \unit{5}{\femto\second}.}
\label{fig:serietempo}
\end{figure}

\begin{figure}[h!]
\centerline{\includegraphics[width=.8\columnwidth]{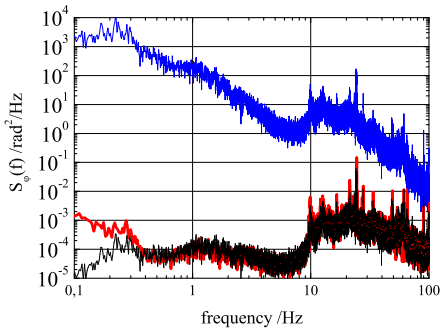}}
\caption{The phase noise spectral density of the one-way signal (blue line), of the phase difference (red line) and the expected limitation due to uncorrelated noise (black line).}
\label{fig:spettri}
\end{figure}

\begin{figure}[h!]
\centerline{\includegraphics[width=.8\columnwidth]{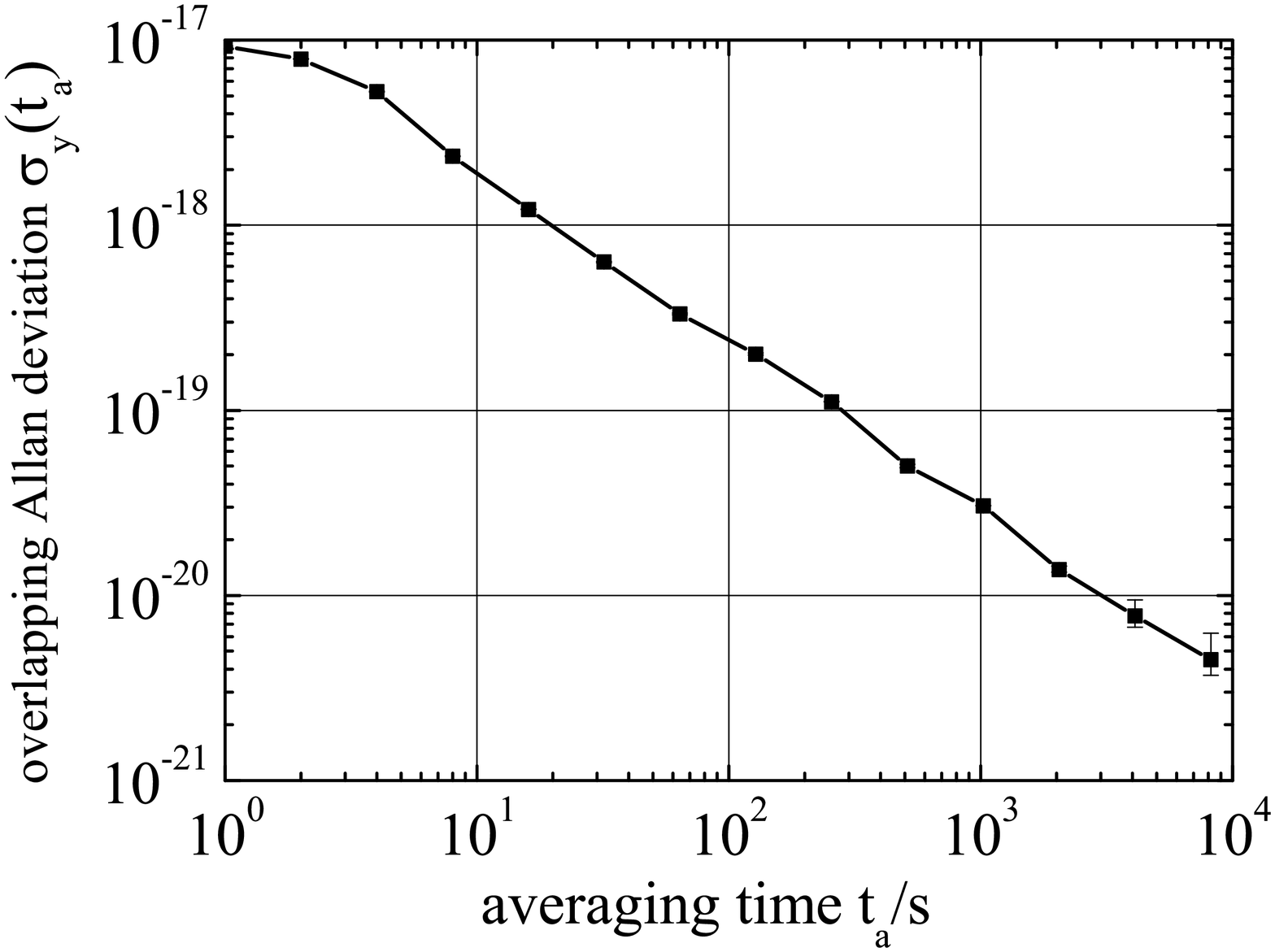}}
\caption{The overlapping Allan deviation \( \sigma_\text{y}(t_\text{a}) \) for the frequency difference on a bandwidth of \unit{0.5}{\hertz}.}
\label{fig:stabilita}
\end{figure}

\begin{figure}[h!]
\centerline{\includegraphics[width=.8\columnwidth]{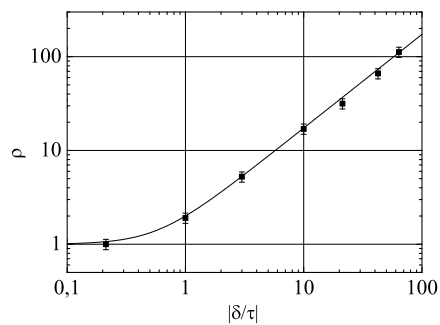}}
\caption{Measured (squares) and calculated (line) degradation of the phase noise when an additional delay \(\delta\) due to a bad synchronization is included. The graph shows \( \rho=\sqrt{S_{\varphi,\text{D}} (f) / S^0_{\varphi,\text{D}} (f)} \) at \(f=\) \unit{1}{\hertz}, where \( S_{\varphi,\text{D}} (f) \) and \(S^0_{\varphi,\text{D}} (f) \) are the PSD of delayed and synchronously subtracted phase samples, as a function of  \( \vert \delta / \tau \vert \).}
\label{fig:spettriShift}
\end{figure}

The advent of optical clocks has enabled frequency metrology to achieve the \( 10^{-18} \) level of uncertainty \cite{chou}. These extremely high performances pave the way for a number of applications in fundamental physics, high resolution spectroscopy \cite{roseband} and relativistic geodesy \cite{chou2}, but at the same time, they require an adequate technique to perform frequency comparisons between distant clocks. Phase-compensated optical links have proved to be reliable from this point of view and outperform state-of-the-art satellite techniques by orders of magnitude \cite{TWFT}. The trasmission of RF and microwaves \cite{sliv, lopez1, fujieda, he,wang}, optical frequencies \cite{pred,lopez2,williams,levi}, and of  an optical comb \cite{marra} have been demonstrated, and time dissemination has recently been performed as well \cite{lopez3,sliv2,rost}. In coherent optical links the phase noise added by the fiber due to environmental noise is actively cancelled. This is obtained by delivering an optical signal to the remote end and by reflecting a part of the transmitted radiation  back to the local laboratory. Here the round trip signal is compared to the original one, and the phase noise added by a double pass in the fiber is detected and compensated with a phase locked loop (PLL).
Active noise cancellation allows the delivery of  an optical frequency over hundreds of kilometers, with a stability at the \( 10^{-20} \) level. \\
The bridging of long distances poses several issues: the beatnote between the local and the round-trip signal is often deteriorated by undesired backreflections, optical and electrical wideband noise, and is detected with a poor signal to noise ratio (SNR). In addition, amplitude modulation may occur, especially if optical amplifiers are used along the way. In most cases,  a clean-up tracking oscillator is required, to filter the wideband noise and eliminate amplitude modulation. However, if the SNR at detection is low,  the clean-up oscillator is affected by cycle slips, which  result in glitches and possible frequency biases on the delivered signal \cite{ascheid}.  Cycle slips may also happen with non-stationary noise events, or if the tracking oscillator bandwidth is too low.\\
In this work, we investigate an alternative technique for comparing distant ultrastable lasers that does not require the active fiber noise cancellation. The noise is cancelled by data post-processing, in analogy to  two-way methods, such as satellite links for frequency transfer \cite{TWFT}.  Two lasers, with a coherence length longer than the fiber haul, are injected in the link at the two opposite ends and travel along the fiber. Their optical phases are measured at the other end against the local laser. If the same fiber is used in both directions, and the phase measurement is synchronous at the two ends, the link noise is  cancelled when comparing the two datasets. With this technique, the beatnotes are less sensitive to optical losses, noise, and backreflections, thanks to the fact that light travels only once in the fiber. The main requirement  is the synchronous phase-comparison: in our system, two Tracking Direct Digital Synthesizers (Tracking DDSs) \cite{calosso2} measure the optical phases at each fiber end with negligible delay and no dead time \cite{micalizio,calosso1}. \\
Digital implementation is reliable and can be upgraded to perform other tasks such as time dissemination, with  reduced costs and easier replication than modem-based systems.  Furthermore, this setup may be useful for novel applications of fiber links, such as the investigation of non reciprocal effects in large fiber loops \cite{clivati2}. \\
This Letter describes the optical and the electronic systems. Then, it reports on the results, highlighting advantages  and limitations of this technique.\\
In this work we consider the comparison of two ultrastable lasers at \unit{194}{\tera\hertz} separated by a \unit{47}{\kilo\metre} fiber, that is part of the metropolitan fiber network. This fiber is used for the Internet data traffic and is implemented on a Dense Wavelength Division Multiplexed (DWDM) architecture, with $\sim$\unit{23}{\deci\bel} of optical losses. The 44th channel of the International Telecommunication Union (ITU) grid (wavelength \unit{1542.14}{\nano\metre}) has been dedicated to our experiment, while Internet data are transmitted on the 21st and 22nd channel, \unit{2}{\tera\hertz} away. The fiber has both ends in our laboratory, and  the same laser was used in the two directions, to investigate the ultimate stability of this scheme. 
The  setup is sketched in Figure \ref{fig:setup}. The ultrastable frequency signal at \unit{194}{\tera\hertz} was provided by a fiber laser frequency locked with the Pound-Drever-Hall technique to a Fabry-Perot high-finesse cavity (120,000) made of Corning Ultra Low Expansion (ULE) glass. The resulting laser linewidth is about \unit{30}{\hertz}  \cite{clivati}. The laser was split into two parts that simulated two different lasers located in distant laboratories. At each side, part of the radiation was injected into the fiber, while the remaining radiation served as a local oscillator. We used two Acousto-Optic Modulators (AOMs) at about \unit{40}{\mega\hertz} frequency-separated by nearly \unit{500}{\kilo}{\hertz}, to distinguish the signal coming from the far fiber end from the stray reflections. Two Optical Add\&Drop Multiplexers were used to inject and extract our signal from the multiplexed fiber network.  At each side, the beatnote between the local and the received light was detected with a photodiode, filtered and amplified; then, its phase $\varphi_\text{a}$ ($\varphi_\text{b}$)
was tracked and measured with a system based on a DDS. The phase discriminator is a double balanced mixer. Its output is digitalized through an Analog to Digital Converter (ADC) and fed to a servo, that calculates the correction for the  DDS. Within the PLL bandwidth, the sequence of data sent to the DDS coincides with the tracking phase. This data stream gives direct access to the beatnote phase, without the need for additional instrumentation such as Fast Fourier Transform Spectrum Analyzers, or phase/frequency meters. A Field Programmable Gate Array (FPGA) implements the PLL controllers and guarantees the synchronization at the $\micro$s level. Also, it averages data, thus unambiguously setting the measurement bandwidth. In this implementation the DDS does not act only as a filter, as in classical schemes that rely on VCOs \cite{rubiola},  but as a phase-measurement unit as well. In addition, DDSs have a wide output frequency range, providing additional flexibility to the experiment.  \\ 
The noise and quantization of the DDS and the ADC are negligible in this kind of application. This is because we deal with optical frequencies, in which the typical phase noise is much higher than any contribution from the electronics. This system has a tracking bandwidth of about \unit{20}{\kilo\hertz}, limited by the serial driving of the DDS. This is the minimum bandwidth required by this link, as demonstrated by the presence of some cycle slips (about 10 per hour). By parallel driving the DDS, a bandwidth of up to \unit{1}{\mega\hertz} is feasible, that is suitable for hauls of hundreds kilometers. 
In practice, the bandwidth of the DDS-based PLL must be such that the phase error at closed loop is  minimized,  to prevent cycle-slips \cite{ascheid}. Hence, it must be adequate to track the fiber acoustic noise but, at the same time, the phase noise floor must be sufficiently low. Thus, the real limitation to the tracking bandwidth is  the SNR at detection. In this sense, as is explained below, the two-way scheme is advantageous, allowing full benefit of the  \unit{1}{\mega\hertz} bandwidth.
Figure \ref{fig:serietempo} shows  the time evolution of the beatnote phases as measured at the two fiber ends, and their difference. The fiber accumulated about \unit{60}{\pico\second} in \unit{50000}{\second}, and noise was cancelled at the \unit{0.1}{\femto\second} level when calculating the difference. The initial transitient and the residual noise of the phase difference were due to slow temperature changes of the laboratory, that affected the short, non common optical fibers of the interferometer. The glitches appearing on the phase difference were due to occasional cycle slips. They were not an issue and  have been removed off-line. \\
Figure \ref{fig:spettri} shows the phase noise spectral density (PSD) of one of the two beatnotes (blue circles) and of the phase difference (red line). At Fourier frequencies \(f>\)\unit{1}{\hertz}, the  noise was due to optical length variations uncorrelated with position; the graph shows their expected contribution (black line). At low  frequencies the noise was  dominated by  the short fibers that were not common in the two systems. Their contribution has been measured by replacing the \unit{47}{\kilo\metre} fiber with a \unit{1}{\metre} fiber, and is compatible with the observed behaviour at low frequencies.  \\
The contribution from the noise uncorrelated with position, integrated over the whole fiber, can be evaluated considering the phase variation  \(\delta \varphi_\text{F}(z,t) \) in each fiber segment as a function of position \(z\) and time \(t\). Since two counterpropagating beams travel  along each position at different times, their phase difference \( \delta \varphi_{\text{D}} \) at the output is: 
$$
\delta \varphi_\text{D}(t)=  \int_0^L \delta \varphi_\text{F} (z, t-\tau +n \frac{z}{c}) \, dz- \int_0^L \delta \varphi_ \text{F} (z, t- n\frac{z}{c})\, dz 
$$
where \(\tau=n \frac{L}{c} \) and \(n\) is the refractive index of the fiber.
After performing the Fourier transform of the autocorrelation function and integration, the PSD \( S_{\varphi,\text{D}}(f)\) of the phase difference can be computed:
\begin{equation}
\label{eq:limit}
S_{\varphi,\text{D}}(f)= \frac{1}{3}(2 \pi \tau f)^2 S_{\varphi, \text{F}}(f)
\end{equation}
where \( S_{\varphi, \text{F}}(f) \) is the PSD of the one-way fiber noise, and it has been assumed that the noise PSD does not depend on $z$ \cite{williams}. Eq. \ref{eq:limit} holds in the spectral region where \( 2 \pi f \tau \ll 1 \). The expected contribution, shown in Figure \ref{fig:spettri},  is in agreement with the measurements.\\
In real links, most of the noise is uncorrelated with position. However, it is interesting to note that in principle, for those applications in which the noise is correlated, the noise limitation shown in eq. \ref{eq:limit} could be overcome. \\
Phase data can be differentiated to obtain the instantaneous  beatnote frequency on both link ends. Phase data have been measured with an integration time of \unit{1}{\second}, that corresponds to a measurement bandwidth of  \unit{0.5}{\hertz}.
The stability of the frequency difference is shown in Figure \ref{fig:stabilita} in terms of Allan deviation  \( \sigma_\text{y}(t_\text{a}) \) as a function of the averaging time \(t_\text{a} \), and achieves   \(4 \times 10^{-21}\) at \unit{10^4}{\second}. The mean frequency difference is   \(< 4 \times 10^{-21}\).\\
It is interesting to estimate the  performance deterioration due to a not perfect synchronization of the samples. Following the same approach used to derive eq. \ref{eq:limit},  the time mismatch can be modelled as an additional delay \( \delta\); \( \tau\) is then replaced by \( \tau+\delta\). After some algebraic manipulation, and again assuming that the noise is uncorrelated with position, one ends up with:
\begin{equation}
S_{\varphi,\text{D}}(f)= \Big ( \frac{1}{3}(2 \pi \tau f)^2 + (2 \pi \delta f)^2 \Big ) S_{\varphi, \text{F}}(f),
\label{eq:limitShift}
\end{equation}
holding in the spectral region where \( 2 \pi f ( \tau + \delta) \ll 1 \). A factor \( 3 \big (\frac{\delta}{\tau} \big )^2 \) deterioration is expected with respect to the optimal case, if a delay  $\vert \delta \vert> \tau$ is introduced.\\
This model was confirmed by the experimental data. 
We evaluated the phase noise increase of the phase difference for several values of \( \delta\). Figure \ref{fig:spettriShift} shows the value of the quantity 
\( \rho=\sqrt{S_{\varphi,\text{D}} (f) / S^0_{\varphi,\text{D}} (f)} \)
at \( 	f=\)\unit{1}{\hertz}, where \(S_{\varphi,\text{D}} (f)\) denotes the PSD of the phase difference 
with delayed samples, and \(S^0_{\varphi ,\text{D}} (f)\)  denotes the PSD of the phase difference with synchronously subtracted samples. The graph shows the obtained points (black squares) and the calculated value (line), according to eq. \ref{eq:limitShift}, as a function of \( \vert \delta / \tau \vert \). \\
In practice,  timing at the \(\micro\)s level is feasible, and does not require continuous monitoring, as typical delay variations are negligible at this level \cite{lopez3}. Synchronization at the \(\micro\)s level is widely enough even for a short \unit{47}{\kilo\metre} link (in which $\tau=$\unit{235}{\micro\second}), with a noise increase below 1\%. Timing becomes less stringent for longer links. If data postprocessing is used, an algorithm  can also be developed to minimize \( S_{\varphi,\text{D}} (f) \), avoiding the need for a precise  synchronization. This may be helpful if some segments of the link are much noisier than others, as the algorithm can be optimized to cancel their contribution.\\ 
In summary, we implemented a two-way optical frequency transfer technique on optical fiber, based on the remote synchronous measurement of the optical phase, and demonstrated its performance at the \(10^{-21}\) level of stability. This scheme may be useful when a clock comparison  and no frequency dissemination is needed. The difference of the two frequencies is at first order insensitive to the fiber noise,  as two beams counterpropagate in the same fiber. Some technical limitations of actively compensated fiber links become less stringent, as in this setup each beam travels only once in the fiber. So, the optical carrier is affected by half phase noise, by a lower wideband noise of optical amplifiers, and by less amplitude modulation; optical power and SNR at the two ends are higher, thus enabling a higher tracking bandwidth and possibly to use less amplifiers. These aspects are  especially desirable in long optical links, and allow a better signal tracking and  significantly less cycle slips. The digital architecture allows fast tuning, and most system upgrades are feasible just with additional firmware. Moreover, the data of interest, such as phase, frequency, signal power, are routinely measured inside the FPGA, and can be monitored without any external instrumentation. Thus, this scheme is suitable for autonomous and remotely controlled link operation.\\

The authors thank E. Rubiola and G. Santarelli for useful discussions, and the GARR Consortium for technical help with the fibers. \\
This  work  was  partly funded  by  Compagnia  di  San  Paolo,  by MIUR under Progetto Premiale 2012, and  by the  EMRP  program 
(SIB02-NEAT-FT).  The  EMRP  is  jointly  funded  by  the  EMRP  participating countries within EURAMET and the European Union.\\

\end{document}